# Comments on "Nonrenewal Statistics in the Catalytic Activity of Enzyme Molecules at Mesoscopic Concentrations"

In-Chun Jeong,[1-3] Sanggeun Song,[1-3] Daehyun Kim,[1-3] Seong Jun Park,[1-3] Ji-Hyun Kim[1] and Jaeyoung Sung[1-3*]

[1]*Creative Research Initiative Center for Chemical Dynamics in Living Cells, Chung-Ang University, Seoul 06974, Korea.*

[2]*Department of Chemistry, Chung-Ang University, Seoul 06974, Korea.*

[3]*National Institute of Innovative Functional Imaging, Chung-Ang University, Seoul 06974, Korea.*

[*]jaeyoung@cau.ac.kr

A superposition of independent, renewal processes forms a nonrenewal process with serial correlations among time intervals between successive events [1,2], which is confirmed for enzyme processes in Ref. [3]. The authors of Ref. [3] further claimed that the mean catalytic rate of a system of enzymes at mesoscopic concentration does not obey the Michaelis-Menten (MM) equation, even if the MM equation is correct for a single enzyme as well as for a macroscopic number of enzymes [4-6]. However, in fact, the MM equation provides the correct steady-state rate for the model in Ref. [3] regardless of the number of enzymes. Using the relationship between the mean and characteristic function,

$$\frac{d\langle N_P(t)\rangle_{N_E}}{dt} = \frac{d}{dt}\left[\frac{\partial}{\partial s_2}G(s_1=1,s_2,t)\bigg|_{s_2=1}\right],$$

and the characteristic function, $G(s_1,s_2,t)$, given in Eq. (4) of Ref. [3], one easily recovers the MM equation in the steady-state.

It is well known in enzyme kinetics that the MM equation holds for enzymes in the steady-state. The authors of Ref. [3] obtained the result inconsistent with the MM equation because they investigated the enzyme system in a nonstationary state. For example, Eq. (7) in Ref. [3] is the probability density of the time elapsed for the very first enzymatic turnover in $N$ enzyme system under the synchronized initial condition (Fig. 1(a)), under which all the enzymes in the system synchronously start catalytic reactions at time 0. The authors showed that the first moment of Eq. (7) in Ref. [3] does not satisfy the MM equation, which is expected because the equation is not the enzymatic turnover time distribution of $N$-enzyme system in the steady-state.

The correct turnover time distribution, $\psi_N^{st}(t)$, for a system of N enzymes in the steady-state, shown in Fig. 1(b), is given by $\psi_N^{st}(t) = -\partial S_N(t)/\partial t$ where $S_N(t)$ is the probability that none of enzymes in the system has completed a catalytic turnover during time $t$, since an enzyme, say $E_1$, in the system completed the last catalytic turnover event at time 0. In the steady-state, $S_N(t)$ can be factorized into $S_N(t) = S_1(t)\left[S_1^{st}(t)\right]^{N-1}$ where $S_1(t)$ and $[S_1^{st}(t)]^{N-1}$ denote, respectively, the probability that enzyme $E_1$ has not completed another catalytic turnover till time $t$ and the probability that none of the remaining $N$-1 enzymes has completed a catalytic turnover event till time $t$, since $E_1$ completed the last catalytic turnover at time 0. $S_1(t)$ and $S_1^{st}(t)$ are given by $S_1(t) = \int_t^\infty d\tau \psi_1(\tau)$ and $S_1^{st}(t) = \int_t^\infty d\tau \psi_1^{st}(\tau)$, where $\psi_1(t)$ and $\psi_1^{st}(t)$ denote the turnover time distribution of a single enzyme and $\psi_1^{st}(t) = S_1(t)\Big/\int_0^\infty dt S_1(t)$, respectively [7,8]. For the enzyme model considered in Ref. [3], $\psi_1(t)$ is given by Eq. (3) in Ref. [9]. From these equations, we obtain

$$\psi_N^{st}(t) = \psi_1(t)\left[S_1^{st}(t)\right]^{N-1} + (N-1)\psi_1^{st}(t)S_1(t)\left[S_1^{st}(t)\right]^{N-2}. \tag{1}$$

The first moment $\langle t \rangle_N^{st} \left(\equiv \int_0^\infty dt\, t\, \psi_N^{st}(t)\right)$ of $\psi_N^{st}(t)$ obeys the MM equation, if the mean enzymatic turnover time $\langle t \rangle_1$ of a single enzyme does [5,6], because $\langle t \rangle_N^{st} = \int_0^\infty dt\, S_N(t) = -\langle t \rangle_1 \int_0^\infty dt\, \frac{dS_1^{st}(t)}{dt}\left[S_1^{st}(t)\right]^{N-1} = \langle t \rangle_1 / N$, which is in direct contradiction with Ref. [3].

In Fig. 2, a comparison between our theory and simulation is made, which clearly shows that Eq. (1) is the correct turnover time distribution of the system of $N$ enzymes in the steady-state; in contrast, Eq. (7) in Ref. [3] is the probability density, $\psi_N^{sync}(t|1)$, of the *first* turnover time in the system of $N$ enzymes under the synchronized initial condition. Under the synchronized initial condition, the enzymatic turnover time distribution, $\psi_N^{sync}(t|i)$, of the $i$-th enzymatic turnover time, $t_i$, depends on the number, $i$, of the enzymatic turnover event, but in the large $i$ limit, $\psi_N^{sync}(t|i)$ approaches to $\psi_N^{st}(t)$; therefore, the mean enzymatic turnover time under the synchronized condition approaches to $\langle t \rangle_N^{st}$ in the large $i$ limit, which satisfies the MM equation.

**Acknowledgment**

This research was supported by the Creative Research Initiative Project program (No. 20151001) funded by National Research Foundation (NRF) of Korea government. I.-C. Jeong gratefully acknowledges the Chung-Ang University Research Scholarship Grants in 2015.

**Figures**

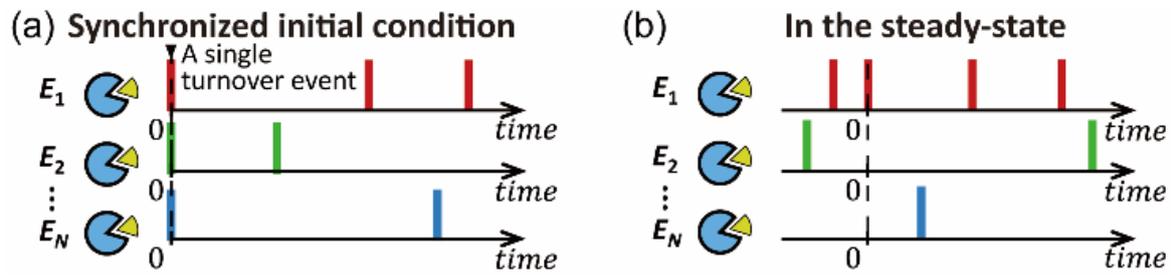

FIG. 1. A system of $N$ enzymes (a) under the synchronized initial condition (b) in the steady state

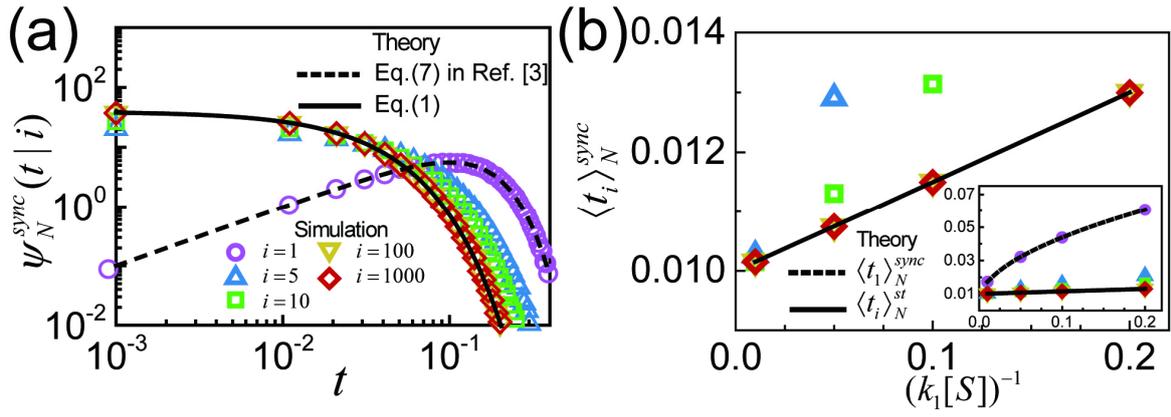

FIG. 2. Comparison between theory and simulation. (a) $\psi_N^{sync}(t|i)$ (b) $\langle t_i \rangle_N^{sync} \left[ = \int_0^\infty dt \psi_N^{sync}(t|i) t \right]$. $N=100$. The values of other parameters are the same as those used in Ref. [3].